\documentclass[journal]{IEEEtran}
\usepackage[T1]{fontenc}
\usepackage[utf8]{inputenc}
\usepackage{amsmath}
\usepackage{amsthm}
\usepackage{amssymb}
\usepackage{graphicx}
\allowdisplaybreaks
 \usepackage[letterpaper,top=0.5in, bottom=0.5in, left=0.5in, right=0.5in]{geometry} 
 \makeatother
\makeatletter
\theoremstyle{plain}
\newtheorem{thm}{\protect\theoremname}
\theoremstyle{plain}
\newtheorem{cor}[thm]{\protect\corollaryname}

\usepackage{balance}
\usepackage{cite}
\usepackage{acronym}
\usepackage{graphicx}
\usepackage{comment}
\AtBeginDocument{\acrodef{6G}{sixth-generation}
\acrodef{ALM}{augmented Lagrangian method}
\acrodef{AO}{alternating optimization}
\acrodef{AoA}{angle of arrival}
\acrodef{AoD}{angle of departure}
\acrodef{APGM}{alternating projected gradient method}
\acrodef{APM}{accelerated proximal gradient method}
\acrodef{AP}{access point}
\acrodef{ASP}{antenna separation product}
\acrodef{AWGN}{additive white Gaussian noise}
\acrodef{BC}{broadcast channel}
\acrodef{BCM}{block coordinate maximization}
\acrodef{BEP}{bit error probability}
\acrodef{BER}{bit error rate}
\acrodef{BF-MIMO}[BF\mbox{-}MIMO]{beamforming MIMO}
\acrodef{BF}{beamforming}
\acrodef{BS}{base station}
\acrodef{bpcu}{bits per channel use}
\acrodef{CNN}{convolutional neural networks}
\acrodef{CP}{cyclic prefix}
\acrodef{CPU}{central processing unit}
\acrodef{CR}{communication rate}
\acrodef{CSI}{channel state information}
\acrodef{CSIR}{channel state information at RX}
\acrodef{SSK}{space shift keying}
\acrodef{CRLB}{Cram\'er-Rao lower bound}
\acrodef{CSIT}{channel state information at TX}
\acrodef{DCMC}{discrete\mbox{-}input continuous\mbox{-}output memoryless channel}
\acrodef{DFT}{discrete Fourier transform}
\acrodef{DL}{deep learning}
\acrodef{DL-TR-GSM}{dual-layered transmit-receive \acl{GSM}}
\acrodef{DLT}{dual-layered transmission}
\acrodef{DMA}{dynamic metasurface antenna}
\acrodef{DOA}{direction of arrival}
\acrodef{DoF}{degrees of freedom}
\acrodef{DNN}{deep neural network}
\acrodef{DPC}{dirty paper coding}
\acrodef{DRL}{deep reinforcement learning}
\acrodef{EE}{energy efficiency}
\acrodef{EGC}{equal gain combining}
\acrodef{EM}{electromagnetic}
\acrodef{EVD}{eigenvalue decomposition}
\acrodef{FPGA}{field programmable gate array}
\acrodef{FSPL}{free space path loss}
\acrodef{FFT}{fast Fourier transform}
\acrodef{FDE}{frequency domain equalization}
\acrodef{GRSM}{generalized \acl{RSM}}
\acrodef{GSM}{generalized \acl{SM}}
\acrodef{HMIMO}{holographic MIMO}
\acrodef{IA}{inner approximation}
\acrodef{IFFT}{invserse fast Fourier transform}
\acrodef{ICI}{inter-channel interference}
\acrodef{iid}[i.i.d.]{independent and identically distributed}
\acrodef{IMT}{International Mobile Telecommunications}
\acrodef{IQ}{in\mbox{-}phase and quadrature}
\acrodef{ISAC}{integrated sensing and communication}
\acrodef{ISI}{intersymbol interference}
\acrodef{ISI-free}[ISI\mbox{-}free]{intersymbol interference free}
\acrodef{LIS}{large intelligent surface}
\acrodef{LOS}{line\mbox{-}of\mbox{-}sight}
\acrodef{KKT}{Karush\mbox{-}Kuhn\mbox{-}Tucker} 
\acrodef{MA}{movable antenna}
\acrodef{MAC}{multiple-access channel}
\acrodef{mmWave}{millimeter-wave}
\acrodef{MI}{mutual information}
\acrodef{MIMO}{multiple\mbox{-}input multiple\mbox{-}output}
\acrodef{mMIMO}{massive MIMO}
\acrodef{MISO}{multiple\mbox{-}input single\mbox{-}output}
\acrodef{ML}{maximum likelihood}
\acrodef{MRC}{maximal ratio combining}
\acrodef{MMSE}{minimum mean square error}
\acrodef{MU-TR-GSM}{multiuser transmit-receive  \acl{GSM} }
\acrodef{NCSIT}{no channel state information at TX}
\acrodef{NLOS}{non\mbox{-}\acs{LOS}} 
\acrodef{NOMA}{non-orthogonal multiple access}
\acrodef{OFDM}{orthogonal frequency division multiplexing}
\acrodef{OFDMA}{orthogonal frequency division multiple access}
\acrodef{OMP}{orthogonal matching pursuit}
\acrodef{OTFS}{orthogonal time frequency space}
\acrodef{UAV}{unmanned aerial vehicle}
\acrodef{umMIMO}{ultra-massive MIMO}
\acrodef{PA}{pinching antenna}
\acrodef{PASS}{pinching antenna system}
\acrodef{PAE}{power added efficiency}
\acrodef{PAPR}{peak\mbox{-}to\mbox{-}average power ratio}
\acrodef{PDF}{probability density function}
\acrodef{PEP}{pairwise error probability}
\acrodefplural{PEP}{pairwise error probabilities}
\acrodef{PGM}{projected gradient method}
\acrodef{PMP}{probability mass function}
\acrodef{PSM}{precoding-aided spatial modulation}
\acrodef{QSM}{quadrature spatial modulation}
\acrodef{RC}{reorganization computation}
\acrodef{RCS}{radar cross section}
\acrodef{RF}{radio frequency}
\acrodef{RHS}{right-hand side}
\acrodef{RIS}{reconfigurable intelligent surface}
\acrodef{RSM}{receive spatial modulation}
\acrodef{RX}{receiver}
\acrodef{SDR}{semi-definite relaxation}
\acrodef{SE}{spectral efficiency}
\acrodef{SEP}{symbol error probability}
\acrodef{SER}{symbol error rate}
\acrodef{SIC}{successive interference cancellation}
\acrodef{SIM}{stacked intelligent metasurface}
\acrodef{SINR}{signal-to-interference-plus-noise ratio}
\acrodef{SISO}{single-input single-output}
\acrodef{SM}{spatial modulation}
\acrodef{SMX-MIMO}[SMX\mbox{-}MIMO]{spatial multiplexing MIMO}
\acrodef{SMX}{spatial multiplexing}
\acrodef{SNR}{signal-to-noise ratio}
\acrodef{SC}{single carrier}
\acrodef{SCA}{successive convex approximation}
\acrodef{SVD}{singular value decomposition}
\acrodef{SPST}{single pole single-throw}
\acrodef{SR}{sensing rate}
\acrodef{SU}{secondary user}
\acrodef{TDE}{time domain equalization}
\acrodef{THz}{terraherz}
\acrodef{TX}{transmitter}
\acrodef{ULA}{uniform linear array}
\acrodef{URA}{uniform rectangular array}
\acrodef{VGA}{variable gain amplifier}
\acrodef{WSR}{weighted sum rate}
\acrodef{wrt}[w.r.t.]{with respect to}
\acrodef{ZF}{zero-forcing}
\acrodef{ZMCG}{zero-mean complex Gaussian}

}
\makeatother
\providecommand{\corollaryname}{Corollary}
\providecommand{\theoremname}{Theorem}
\usepackage[font=small,labelfont=small]{caption}

\begin{document}
\title{DL-Driven Optimization for ISAC System Equipped With Pinching and Movable Antennas}

\author{Nemanja Stefan Perovi\'c,~\IEEEmembership{Member,~IEEE}, Keshav Singh,~\IEEEmembership{Senior Member, IEEE}, and Chih-Peng Li,~\IEEEmembership{Fellow, IEEE}

\thanks{N. S. Perovi\'c, K. Singh, and C.-P. Li are affiliated with the Institute of Communications Engineering, National Sun Yat Sen University, Kaohsiung 80424, Taiwan (e-mail:  \{n.s.perovic,\,keshav.singh,\,cpli\}@mail.nsysu.edu.tw).} 


\vspace{-2.25em}
}

\maketitle
\begin{abstract}
\Ac{ISAC} is considered one of the key technologies for future wireless networks, due to its ability to provide accurate environment awareness and improve energy, spectral and hardware efficiency.
Moreover, the introduction of recently developed \acp{PA}
and \acp{MA} has the potential to further improve the performance
gains of \acs{ISAC}. Therefore, our goal is to study the optimization of the sum-rate for
an ISAC system equipped with PAs and MAs, capable of satisfying minimal sensing requirements. To achieve it, we derive a closed-form
solution for the optimal sensing receive combiner, and show that it
is determined by other optimization variables. For these other variables
(i.e., the positions of the transmit PAs, the positions of the users'
MAs, the communication precoding matrices, and the sensing transmit
beamformer), we propose a \ac{DL} network to find their optimal
values. To train the network in an unsupervised manner, we formulate
a loss function consisting of the objective function, as well as
the penalty terms related to the constraints for the PAs and MAs positions.
Simulation results show that using PAs and MAs in ISAC systems provides
a larger sum-rate compared to ISAC systems with only fixed antennas,
and that this performance advantage is increased with the maximum
transmit power. Furthermore, we demonstrate that the communication
performance of the considered system is a bit more affected by the
sensing \ac{SINR} threshold compared to the sensing performance.\acresetall{}
\end{abstract}
\vspace{-0.25em}
\vspace{-2.4em}

\section{Introduction}
\bstctlcite{BSTcontrol}

\Ac{ISAC}, which merges radar-style sensing and data
transmission through the co-design of shared waveforms, spectrum, and hardware resources, is
regarded as one of the key technologies for 6G and beyond networks due to the potential to enable
many new applications with accurate environment awareness, such as vehicular networks, industrial
robotics and automation, intelligent interactive networks.
\Ac{mMIMO} remains a foundational technology for ISAC
in 6G systems with an increasing number of antennas for achieving high spatial multiplexing,
spatial resolution and directional beamforming. These large-scale antenna arrays that allow the system to
serve multiple users simultaneously, provide higher channel capacity for data transmissions and more
accurate sensing. However, besides facing increased hardware complexity and energy consumption, due to fixed-position antennas,  conventional \ac{mMIMO} cannot fully exploit the spatial variation of wireless channels nor adapt the optimal array geometries for sensing and communication applications.

To overcome these limitations, a new antenna architecture with \acp{MA} has been proposed for achieving additional 
\acp{DoF} in the spatial domain. By using electromechanical mechanisms, the positions of MAs can be flexibly adjusted within a designated spatial area whose size typically spans from several to tens of wavelengths, thereby reconstructing channel conditions to boost communication performance or reconfiguring the geometric properties of MIMO arrays to enhance sensing capability.  
In \cite{ma2024movable}, the \ac{CRLB} for \ac{AoA} estimation was derived as a function of the MAs' positions and later
optimized. An optimization framework for maximizing the sensing \ac{SINR}
in a multi-user bistatic ISAC system, where MAs were deployed for adjusting
the array responses at both the transmit and the receive \acp{BS},
was introduced in \cite{jiang2025movable}. The maximization of the
weighted sum of the communication and the sensing rates in a full-duplex
ISAC communication system, where a single BS was equipped with the
transmit and the receive MAs was considered in \cite{ding2025movable}.
Similarly, the authors in \cite{guo2025movable} studied the joint
active beamforming and MAs' positions optimization for a full-duplex
ISAC network with multiple transmit and receive \acp{BS} that
provided significant communication performance improvements, while
maintaining the guaranteed sensing requirements.

Recently, \acp{PA} have emerged as a novel flexible-antenna technology, which employs dielectric waveguides as the transmission medium. \Ac{EM} waves are radiated by attaching small dielectric particles to the waveguide, enabling PA to be independently activated or deactivated at arbitrary positions along the waveguide. Therefore, \acp{PA} are capable of creating strong \ac{LOS} links so that large-scale path losses and \ac{LOS} blockages can be effectively mitigated by placing antennas close to users. In addition to the high reconfigurability of PA systems, whereby both the locations and the number of pinching antennas can be flexibly adjusted, PAs can be implemented in a simple and cost-effective manner.
In \cite{ouyang2026rate}, closed-form
solutions for the optimal pinching antenna location were derived under
sensing-centric, communications-centric, and Pareto-optimal designs.
The sensing \ac{CRLB} for a PA system was derived and later maximized by
using the proposed optimization framework in \cite{li2026pinching}.
In \cite{zhang2025integrated}, the authors introduced an optimization
algorithm for maximizing the sensing target illumination power, which
also satisfied the specified communication quality-of-service requirements.

Despite all the previously mentioned works \cite{ouyang2026rate}, \cite{li2026pinching},  \cite{jiang2025movable} and \cite{guo2025movable}, none of them considered  
systems that are simultaneously equipped with PAs and MAs. 
The only paper studying communication
systems with both PAs and MAs is \cite{kang2025campass}, where a
\ac{DL} optimization framework was proposed for the achievable rate
optimization in a point-to-point communication scenario.
Unlike that, we investigate an ISAC system that combines the flexibility of PAs to move along the waveguides over a length of dozens of meters at the ISAC transmitter and the limited movement of MAs inside regions whose dimensions enable their deployment within user equipment. 
In that way, we can unify the gains of both PA and MA technologies in a single ISAC system.

Against this
background, the contributions of this paper are listed as follows:
\begin{itemize}
\item We consider an ISAC system equipped simultaneously by both MAs and PAs.  Leveraging their different and compatible capabilities, we develop an optimization framework for maximizing the sum-rate of this ISAC system while maintaining minimal sensing requirements.
\item Inspired by the DL optimization approach for a single-user communication system in  \cite{kang2025campass}, we develop a DL optimization framework for a multi-user ISAC system, with added sensing receive combiner and sensing transmit beamformer.

\item We provide a closed-form solution for the optimal
sensing receive combiner and show that it is determined by the values
of other optimization variables. Next, the optimal values of these variables
(i.e., the positions of the transmit PAs, the positions of the users'
MAs, the communication precoding matrices, and the sensing transmit
beamformer) are obtained by the proposed DL network.  To train this
network, we formulate a loss function that consists of the objective
function, as well as the penalty terms related to the constraints
for the PAs' and MAs' positions.
\item Simulation results show that the proposed scheme achieves improved
performance compared to a benchmark scheme equipped with only fixed
antennas and that this performance advantage increases with the maximum transmit
power. Also, we demonstrate that the change
of the sensing SINR threshold influences the communication performance slightly more than the sensing performance. 
\end{itemize}

\vspace{-1.25em}
\section{System Model}

We consider an ISAC system, in which the PA transmitter is implemented
by $N_{t}$ parallel waveguides, where each waveguide has exactly
one PA. It transmits the ISAC signal, which simultaneously provides
communication with $K$ users and detection of a sensing target. The
reflected echo signal from the sensing target is received by $N_{r}$
fixed antennas, which are spatially separated from the PA transmitter.
Every user possesses $N_{k}$ MAs that can be moved inside the region
that is denoted as $\mathcal{C}_{k}$. 
\begin{figure}[t]
\begin{centering}
\includegraphics[totalheight=4.5cm]
{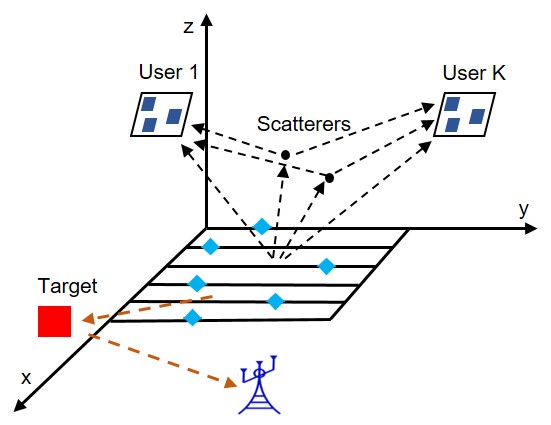}
\par\end{centering}
\caption{The proposed ISAC system with the transmit PAs and
users' MAs.\label{fig:System-model}}
\end{figure}

To better illustrate the considered ISAC system, we introduce 3D coordinate
system, as shown in Fig. \ref{fig:System-model}. The transmit waveguides
are placed in the $xy$-plane, parallel to the $y$-axis. It is assumed
that all the waveguides have the same length of $D_{t}$ and are uniformly
spaced over the interval $[0,D_{t}]$. Therefore, the position of
the PA that is allocated to the $n$-th waveguide is given by $\mathbf{t}_{n}=[x_{t,n},y_{t,n},0]^{T}$,
where $x_{t,n}=(n-1)D_{t}/(N_{t}-1)$ and $y_{t,n}\in[0,D_{t}]$.
The receive \ac{ULA} of length $D_{r}$ is placed parallel to the
$y$-axis, such that its midpoint is located at $[x_{r},y_{r},z_{r}]$.
In addition, it is assumed that the MAs' regions of all users are parallel
to the $xy$-plane. Therefore, the local coordinates of the $b$-th
MA of user $k$ are specified as $\mathbf{\tilde{u}}_{k,b}=[\tilde{x}_{k,b}^{u},\tilde{y}_{k,b}^{u},0]^{T}$,
where $\tilde{x}_{k,b}^{u},\tilde{y}_{k,b}^{u}\in[0,D_{k}]$. With
respect to the global coordinate system, the coordinates of that same
MA are given by $\mathbf{u}_{k,b}=\mathbf{\tilde{u}}_{k,b}+\mathbf{b}_{k}$,
where $\mathbf{b}_{k}$ denotes the global coordinate of the origin
in the local coordinate system. Moreover, the position of the sensing
target is specified as $\mathbf{q}_{p}=[x^{p},y^{p},z^{p}]^{T}$.

In the considered system, the transmitted ISAC signal can be represented as \vspace{-0.5em}
\begin{equation}
\mathbf{x}=\mathbf{F}\left(\sum\nolimits_{k=1}^{K}\mathbf{W}_{k}\mathbf{s}_{k}+\mathbf{v}s_{p}\right),
\end{equation}
where $\mathbf{W}_{k}\in\mathbb{C}^{N_{t}\times N_{k}}$ is the precoding
matrix for transmission to user $k$, and $\mathbf{s}_{k}\in\mathbb{C}^{N_{k}\times1}$
is the signal vector transmitted to the same user. It is assumed that
$\mathbf{s}_{k}$ consists of \ac{iid} symbols that are distributed
according to the complex Gaussian distribution with zero mean value
and unit variance, i.e., $\mathcal{CN}(0,1)$. In addition, $\mathbf{v}\in\mathbb{C}^{N_{t}\times1}$
is the sensing transmit beamformer and $s_{p}$ is
the corresponding sensing signal that satisfies $\mathbb{E}\{|s_{p}|^{2}\}=1$.
The signal propagation through the transmit waveguides is modeled by the
diagonal matrix $\mathbf{F}$, whose diagonal elements are given by $\mathbf{F}(n,n)=e^{-j2\pi y_{t,n}n_{e}/\lambda}$,
where $\lambda$ is the wavelength and $n_{e}$ is the effective refractive
index of the dielectric waveguide.

\vspace{-1em}
\subsection{Communication Model}

The receive signal vector for user $k$ can be expressed as 
\begin{align}
\!\!\mathbf{y}_{C,k} & =\mathbf{H}_{k}\mathbf{F}\mathbf{W}_{k}\mathbf{s}_{k}\!+\!\mathbf{H}_{k}\mathbf{F}\!\sum_{u=1, u\neq k}^{K}\!\mathbf{W}_{u}\mathbf{s}_{u}\!+\!\mathbf{H}_{k}\mathbf{F}\mathbf{v}s_{p}\!+\!\mathbf{n}_{k},\label{eq:RecSigUserk}
\end{align}
where $\mathbf{H}_{k}\in\mathbb{C}^{N_{k}\times N_{t}}$ is the channel
matrix between the PA transmitter and user $k$, and $\mathbf{n}_{k}\in\mathbb{C}^{N_{k}\times1}$
is the noise vector with i.i.d. elements distributed according to
$\mathcal{CN}(0,\sigma_{k}^{2})$, where $\sigma_{k}^{2}$ is the
noise variance at user $k$. The channel matrix $\mathbf{H}_{k}$ is modeled by
using the Rician distribution and can be expressed as
\begin{equation}
\mathbf{H}_{k}=\sqrt{\kappa/(\kappa+1)}\mathbf{H}_{L,k}+\sqrt{1/(\kappa+1)}\mathbf{H}_{N,k}
\end{equation}
where $\mathbf{H}_{L,k}$ and \textbf{$\mathbf{H}_{N,k}$ }are the
\ac{LOS} and the \ac{NLOS} channel matrices between the transmitter
and user $k$, respectively, and $\kappa$ is the Rician factor. The
LOS channel between the PA of waveguide $n$ and the $b$-th MA of
user $k$  is given as
\begin{equation}
\mathbf{H}_{L,k}(b,n)=\frac{\lambda}{4\pi||\mathbf{t}_{n}-\mathbf{u}_{k,b}||}e^{-\frac{2\pi}{\lambda}||\mathbf{t}_{n}-\mathbf{u}_{k,b}||}.
\end{equation}
Similarly, the NLOS channel between the same PA and the same user's
MA can be expressed as \cite{kang2025campass}
\begin{equation}
\mathbf{H}_{N,k}(b,n)=\frac{\lambda}{4\pi}\sum\nolimits_{i=1}^{L}\frac{e^{-\frac{2\pi}{\lambda}(||\mathbf{u}_{k,b}-\text{\ensuremath{\xi}}_{i}||-||\mathbf{t}_{n}-\text{\ensuremath{\xi}}_{i}||)}}{||\mathbf{t}_{n}-\text{\ensuremath{\xi}}_{i}||||\mathbf{u}_{k,b}-\text{\ensuremath{\xi}}_{i}||},
\end{equation}
where $L$ is the number of scatterers and $\text{\ensuremath{\xi}}_{i}$
is the position of scatterer $i$. 

From \eqref{eq:RecSigUserk}, the achievable rate for user $k$ can
be written as
\begin{gather}
R_{C,k}=\log_{2}\Bigg|\sigma_{k}^{2}\mathbf{I}_{N_k}+\mathbf{H}_{k}\mathbf{F}\mathbf{W}_{k}\mathbf{W}_{k}^{H}\mathbf{F}^{H}\mathbf{H}_{k}^{H}\Bigg(\sigma_{k}^{2}\mathbf{I}_{N_k}+\nonumber \\
\mathbf{H}_{k}\mathbf{F}\Bigg(\sum\nolimits_{u=1, u\neq k}^{K}\mathbf{W}_{u}\mathbf{W}_{u}^{H}+\mathbf{v}\mathbf{v}^{H}\bigg)\mathbf{F}^{H}\mathbf{H}_{k}^{H}\Bigg)^{-1}\Bigg|.\label{eq:Rk}
\end{gather}

\vspace{-0.4 cm}
\subsection{Sensing Model}

The reflected sensing signals at the receive ULA can be expressed
as \vspace{-0.25em}
\begin{equation}
\mathbf{y}_{S}=\mathbf{G}\mathbf{F}\left(\sum\nolimits_{k=1}^{K}\mathbf{W}_{k}\mathbf{s}_{k}+\mathbf{v}s_{p}\right)+\mathbf{z},
\end{equation}
where $\mathbf{z}\in\mathbb{C}^{N_{r}\times1}$ is the noise vector
with i.i.d. elements that are distributed according to $\mathcal{CN}(0,\sigma_{z}^{2})$,  where $\sigma_{z}^{2}$ is the
noise variance at the sensing receiver.
For a point target, the sensing channel matrix is given by 
\begin{equation}
\text{ \ensuremath{\mathbf{G}}}=\mathbf{f}_{r}\mathbf{f}_{t}^{H},
\end{equation}
where $\mathbf{f}_{t}=\frac{\lambda}{4\pi}[e^{-j\frac{2\pi}{\lambda}||\mathbf{t}_{1}-\mathbf{q}_{p}||}/||\mathbf{t}_{1}-\mathbf{q}_{p}||,\dots,e^{-j\frac{2\pi}{\lambda}||\mathbf{t}_{N_{t}}-\mathbf{q}_{p}||}/||\mathbf{t}_{N_{t}}-\mathbf{q}_{p}||]^{T}$
is the channel between the PA transmitter and the target, $\mathbf{f}_{r}=\frac{\lambda}{4\pi}[e^{-j\frac{2\pi}{\lambda}||\mathbf{r}_{1}-\mathbf{q}_{p}||}/||\mathbf{r}_{1}-\mathbf{q}_{p}||,\dots,e^{-j\frac{2\pi}{\lambda}||\mathbf{r}_{N_{r}}-\mathbf{q}_{p}||}/||\mathbf{r}_{N_{r}}-\mathbf{q}_{p}||]^{T}$
is the channel between the target and the receiver \cite{li2026pinching}.

To extract the information from the sensing signal, we implement the
receive sensing combiner $\mathbf{d}$ and obtain
\begin{align}
y_{c,s} & =\mathbf{d}^{H}\mathbf{y}_{S}=\mathbf{d}^{H}\mathbf{G}\mathbf{F}\left(\sum\nolimits_{k=1}^{K}\mathbf{W}_{k}\mathbf{s}_{k}+\mathbf{v}s_{p}\right)+\mathbf{d}_{p}^{H}\mathbf{z}.
\end{align}

From the previous expression, the appropriate sensing SINR can be
written as 
\begin{align}
\gamma_{s} & =P_{s}/{\mathbf{d}^{H}\mathbf{B}\mathbf{d}},\label{eq:SINR}
\end{align}
where $P_{s}=\mathbf{d}^{H}\mathbf{G}\mathbf{F}\mathbf{v}\mathbf{v}^{H}\mathbf{F}^{H}\mathbf{G}^{H}\mathbf{d}$
is the power of the received sensing signal and $\mathbf{B}=\mathbf{G}\mathbf{F}\sum_{k=1}^{K}\mathbf{W}_{k}\mathbf{W}_{k}^{H}\mathbf{F}^{H}\mathbf{G}^{H}+\sigma_{z}^{2}\mathbf{I}_{N_r}$.

\section{Problem Formulation }

Our goal is to maximize the communication sum-rate of the considered
system, while maintaining the requirement that the sensing SINR is
always above the specified level. Therefore, the optimization problem
can be written as:
\begin{subequations}
\label{eq:Opt_prob}
\begin{align}
\max_{\substack{\{y_{t,n}\},\{\tilde{x}_{k,b}^{u}\},\{\tilde{y}_{k,b}^{u}\},\\\{\mathbf{W}_{k}\},\mathbf{v},\mathbf{d}}}
&
\;\sum\nolimits_{k=1}^{K}R_{C,k}\label{eq:obj_fun}\\
\text{s.t.} & \;\text{Tr}\left(\sum\nolimits_{k=1}^{K}\mathbf{W}_{k}\mathbf{W}_{k}^{H}\right)\le P_{\max}\label{eq:Pmax}\\
 & \;\gamma_{s}\ge\gamma_{0},\label{eq:SINRmin}\\
 & \;0\le y_{t,n}\le D_{t},\forall n,\label{eq:PA_space}\\
 & \;||\text{\ensuremath{\mathbf{v}}}||^{2}=1,\label{eq:TX_BF}\\
 & \;||\text{\ensuremath{\mathbf{d}}}||^{2}=1,\label{eq:RX_BF}\\
 & \;0\le\tilde{x}_{k,b}^{u},\tilde{y}_{k,b}^{u}\le D_{k},\forall k,b,\label{eq:region_con}\\
 & \;||\mathbf{\tilde{u}}_{k,b_{1}}-\mathbf{\tilde{u}}_{k,b_{2}}||\ge d_{\min},\label{eq:User_space}\\
 & \;1\le b_{1}\neq b_{2}\le N_{k},\text{\ensuremath{\forall}}k.\nonumber 
\end{align}
\end{subequations}
 Constraint \eqref{eq:Pmax} determines the available transmit power
budget $P_{\max}$ for communication signals. Constraint \eqref{eq:SINRmin}
specifies that the sensing SINR must not go below the SINR threshold
$\gamma_{0}$. Constraints \eqref{eq:PA_space} means that each PA
must be located on a waveguide. Constraints \eqref{eq:TX_BF} and
\eqref{eq:RX_BF} specify that the sensing transmit beamformer and
the sensing receive combiner should be unit power vectors. Finally,
constraints \eqref{eq:region_con} and \eqref{eq:User_space} specify
that the MAs of a single user must be located within the region $\mathcal{C}_{k}$,
while satisfying the minimum inter-MA separation $d_{\min}$. 

\vspace{-0.3cm}
\section{Proposed Solution}

\subsection{Optimization of Sensing Receive Combiner}

The optimal receive sensing combiner $\mathbf{d}$ maximizes the sensing
SINR. Moreover, one can notice that the numerator of \eqref{eq:SINR}
can be expressed as
\[
\mathbf{d}^{H}\mathbf{G}\mathbf{F}\mathbf{v}\mathbf{v}^{H}\mathbf{F}^{H}\mathbf{G}^{H}\mathbf{d}=\mathbf{f}_{t}^{H}\mathbf{F}\mathbf{v}\mathbf{v}^{H}\mathbf{F}^{H}\mathbf{f}_{t}\times\mathbf{d}^{H}\mathbf{f}_{r}\mathbf{f}_{r}^{H}\mathbf{d},
\]
where $\mathbf{f}_{t}^{H}\mathbf{F}\mathbf{v}\mathbf{v}^{H}\mathbf{F}^{H}\mathbf{f}_{t}$
is a nonnegative scalar independent of $\mathbf{d}$. Therefore, the
optimal $\mathbf{d}$ is the solution of the follow problem
\begin{subequations}
\begin{align}
\text{\ensuremath{\underset{\mathbf{d}}{\mathrm{maximize}}}} & \;\frac{\mathbf{d}^{H}\mathbf{f}_{r}\mathbf{f}_{r}^{H}\mathbf{d}}{\mathbf{d}^{H}\mathbf{B}\mathbf{d}}\label{eq:obj_fun-1}\\
\text{s.t.} & \;||\text{\ensuremath{\mathbf{d}}}||^{2}=1,\label{eq:RX_BF-1}
\end{align}
\end{subequations}
and it is given as
\begin{equation}
\text{\ensuremath{\mathbf{d}}}=\mathbf{B}^{-1}\mathbf{f}_{r}/||\mathbf{B}^{-1}\mathbf{f}_{r}||.\label{eq:RX_xomb}
\end{equation}
In other words, the optimal $\text{\ensuremath{\mathbf{d}}}$ does
not need to be optimized directly, as it is determined by the values
of other optimization variables in \eqref{eq:Opt_prob}. We use this
fact in the next subsection, where a neural network for optimizing
these variables is presented.

\vspace{-0.3cm}
\subsection{Optimization of Other Variables}

\begin{figure*}[t]
\centering{}\includegraphics[scale=0.75]{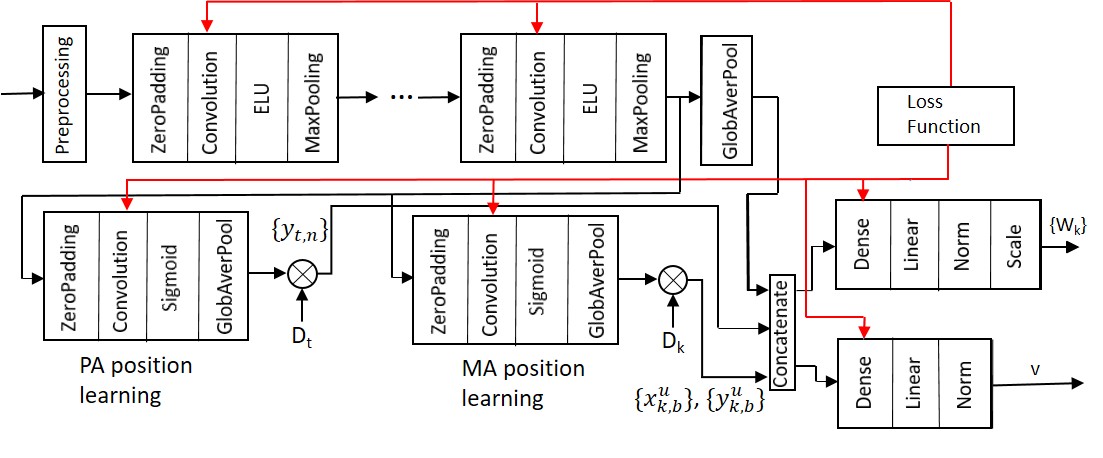}
\vspace{-0.5em}
\caption{Diagram of the considered DL network.\label{fig:Diagram-of-nn}}
\vspace{-1em}
\end{figure*}

Similarly as in \cite{kang2025campass}, the objective function in
\eqref{eq:obj_fun} can be reformulated as
\begin{subequations}
\label{eq:Opt_prob-1}
\begin{align}
\text{\text{\ensuremath{\underset{\{y_{t,n}\},\{\tilde{x}_{k,b}^{u}\},\{\tilde{y}_{k,b}^{u}\}}{\mathrm{max}}}}\;\;\text{\ensuremath{\underset{\mathbf{W}_{k},\mathbf{v}_{p}}{\mathrm{max}}}}} & \;\sum\nolimits_{k=1}^{K}R_{C,k}\label{eq:obj_fun-2}\\
\text{s.t.} & \;\eqref{eq:Pmax},\eqref{eq:SINRmin},\eqref{eq:PA_space},\eqref{eq:TX_BF},\nonumber \\
 & \;\eqref{eq:RX_BF},\eqref{eq:region_con},\eqref{eq:User_space},\nonumber 
\end{align}
\end{subequations}
so that the optimal communication precoders and transmit sensing beamformers
may be determined (learned) from the optimal positions of the PAs
and users' MA. The network architecture that implement this learning
process to solve \eqref{eq:Opt_prob-1} is illustrated in Fig. \ref{fig:Diagram-of-nn},
and is elaborated in more details in the sequel.

\subsubsection{Neural Network Design}

At the beginning first, we perform preprocessing of the input vector
$[\mathbf{y}_{t},\mathbf{b}_{1}^{T},\dots,\mathbf{b}_{K}^{T},\text{\ensuremath{\xi}}_{1}^{T},\dots,\text{\ensuremath{\xi}}_{L}^{T},\mathbf{q}_{p}]$,
where $\mathbf{y}_{t}=[y_{t,1},y_{t,2},\dots,y_{t,N_{t}}]$. More
precisely, every element $e$ of the input vector is normalized by
$(e-e_{\min})/(e_{\max}-e_{\min})$, where $e_{\max}$ and $e_{\min}$
are the maximum and minimum of the feasible set of the element $e$,
respectively.

In the first block, we perform feature extraction from the normalized
input data in order to enable a more efficient learning process. This
is done by using a seven layer \ac{CNN}, since CNNs are well-known
for their feature extraction capabilities. The first layer has $2C$
convolution kernels, while all other consecutive layers are equipped
with $C$ kernels, where $C=N_{t}+2KN_{u}+2KN_{t}N_{k}+2KN_{t}$. In
every layer, the inputs are padded with zeros so that the outputs
have the same size as the inputs. Afterwards, one dimensional convolution
is implemented using convolution kernels of size 3 and a unit stride
rate. Next, the convolution results are going through the exponential
linear unit (ELU), which is an activation function of the convolution
block. Finally, the max-pooling with the pooling window and the stride
rate of 2 is performed.

The outputs of the first block are fed to the second block which predicts
the positions of the transmit PAs and users' MAs. It consists of two convolution
layers with $N_{t}$ and $2KN_{k}$ kernels running in parallel. These
layers have a similar structure as the convolution layers from the
first block, except they used the sigmoid function instead of the
ELU and the global average pooling instead of the max-pooling. Also,
their outputs are scaled with $D_{t}$ and $D_{k}$ to satisfy \eqref{eq:PA_space}
and \eqref{eq:region_con}, respectively.

The third block uses the outputs of the previous two blocks to predict
the communication precoding matrices $\{\mathbf{W}_{k}\}$ and the
sensing beamformer\textbf{ }$\text{\ensuremath{\mathbf{v}}}$. The
outputs the first block pass through global average pooling and are
then combined with outputs of the second block to form the inputs of
the third block. The third block consists of two dense layers, where
the first dense layer predicts all the communication precoding matrices
and the second one predicts the sensing beamformer. The first dense
layer has $2KN_{t}N_{k}$ outputs (i.e., one output for each real/imaginary
part of every precoding coefficient) and is followed by the linear
activation function. These outputs are then normalized to have a unit
norm and later scaled with $\sqrt{P_{\max}}$ to fulfill \eqref{eq:Pmax}.
The other dense layer with $N_{t}$ outputs is also followed by the linear activation function
and the block that normalizes the outputs to satisfy \eqref{eq:TX_BF}. 

\subsubsection{Loss Function}

The training loss function needs to maximize the objective function
\eqref{eq:obj_fun}, while satisfying constraints \eqref{eq:SINRmin}
and \eqref{eq:User_space}. Since neural networks are not designed
to operate with complex numbers, we have to reformulate the user rate
in \eqref{eq:Rk} and the sensing SINR in \eqref{eq:SINR}, so that
they can be calculated using only real numbers. First, the user rate
can be equivalently expressed as 
\begin{align}
R_{C,k}= & \log_{2}\Bigg|\sigma_{k}^{2}\mathbf{I}_{N_k}+\mathbf{H}_{k}\mathbf{F}\Bigg(\sum_{u_1=1}^{K}\mathbf{W}_{u_1}\mathbf{W}_{u_1}^{H}+\mathbf{v}\mathbf{v}^{H}\bigg)\mathbf{F}^{H}\mathbf{H}_{k}^{H}\Bigg|\nonumber \\
- & \log_{2}\Bigg|\sigma_{k}^{2}\mathbf{I}_{N_k}+\mathbf{H}_{k}\mathbf{F}\Bigg(\sum_{u_2=1, u_2\neq k}^{K}\mathbf{W}_{u_2}\mathbf{W}_{u_2}^{H}+\mathbf{v}\mathbf{v}^{H}\bigg)\mathbf{F}^{H}\mathbf{H}_{k}^{H}\Bigg|.
\end{align}
Since the determinant of any positive-definite square matrix $\mathbf{E}$ satisfies the identity
$\left|\mathbf{E}\right|=\sqrt{\left|\begin{array}{cc}
\mathfrak{R}\{\mathbf{E}\} & \mathfrak{I}\{\mathbf{E}^{T}\}\\
\mathfrak{I}\{\mathbf{E}\} & \mathfrak{R}\{\mathbf{E}\}
\end{array}\right|}$, the previous expression can be equivalently written as 
\begin{align}
R_{C,k}= & \frac{1}{2}\log_{2}\Bigg|\begin{array}{cc}
\sigma_{k}^{2}\mathbf{I}_{N_k}+\mathbf{P}_{1} & \mathbf{Q}_{1}^{T}\\
\mathbf{Q}_{1} & \sigma_{k}^{2}\mathbf{I}_{N_k}+\mathbf{P}_{1}
\end{array}\Bigg|\nonumber \\
 & -\frac{1}{2}\log_{2}\Bigg|\begin{array}{cc}
\sigma_{k}^{2}\mathbf{I}_{N_k}+\mathbf{P}_{2} & \mathbf{Q}_{2}^{T}\\
\mathbf{Q}_{2} & \sigma_{k}^{2}\mathbf{I}_{N_k}+\mathbf{P}_{2}
\end{array}\Bigg|,\label{eq:Rk_ref}
\end{align}
where $\mathbf{P}_{1}=\mathfrak{R}\{\mathbf{H}_{k}\mathbf{F}(\sum_{u_1=1}^{K}\mathbf{W}_{u_1}\mathbf{W}_{u_1}^{H}+\mathbf{v}\mathbf{v}^{H})\mathbf{F}^{H}\mathbf{H}_{k}^{H}\}$,
$\mathbf{Q}_{1}=\mathfrak{I}\{\mathbf{H}_{k}\mathbf{F}(\sum_{u_1=1}^{K}\mathbf{W}_{u_1}\mathbf{W}_{u_1}^{H}+\mathbf{v}\mathbf{v}^{H})\mathbf{F}^{H}\mathbf{H}_{k}^{H}\}$,
$\mathbf{P}_{2}=\mathfrak{R}\{\mathbf{H}_{k}\mathbf{F}(\sum_{\underset{u_2\neq k}{u_2=1}}^{K}\mathbf{W}_{u_2}\mathbf{W}_{u_2}^{H}+\mathbf{v}\mathbf{v}^{H})\mathbf{F}^{H}\mathbf{H}_{k}^{H}\}$,
and $\mathbf{Q}_{2}=\mathfrak{I}\{\mathbf{H}_{k}\mathbf{F}(\sum_{\underset{u_2\neq k}{u_2=1}}^{K}\mathbf{W}_{u_2}\mathbf{W}_{u_2}^{H}+\mathbf{v}\mathbf{v}^{H})\mathbf{F}^{H}\mathbf{H}_{k}^{H}\}$.
These matrices can be calculated as
\[
\left[\begin{array}{c}
\mathbf{P}_{1}\\
\mathbf{Q}_{1}
\end{array}\right]=\hat{\mathbf{H}}_{k}\mathbf{\hat{F}}(\sum_{u_1=1}^{K}\mathbf{\hat{W}}_{u_1}\mathbf{\hat{W}}_{u_1}^{T}+\mathbf{\hat{v}}\hat{\mathbf{v}}^{T})\hat{\mathbf{F}}^{T}\left[\begin{array}{c}
\mathfrak{R}\{\mathbf{H}_{k}\}^{T}\\
-\mathfrak{I}\{\mathbf{H}_{k}\}^{T}
\end{array}\right]
\]
\[
\left[\begin{array}{c}
\mathbf{P}_{2}\\
\mathbf{Q}_{2}
\end{array}\right]=\hat{\mathbf{H}}_{k}\mathbf{\hat{F}}(\sum_{u_2=1, u_2\neq k}^{K}\mathbf{\hat{W}}_{u_2}\mathbf{\hat{W}}_{u_2}^{T}+\mathbf{\hat{v}}\hat{\mathbf{v}}^{T})\hat{\mathbf{F}}^{T}\left[\begin{array}{c}
\mathfrak{R}\{\mathbf{H}_{k}\}^{T}\\
-\mathfrak{I}\{\mathbf{H}_{k}\}^{T}
\end{array}\right]
\]
where $\mathbf{\hat{\mathbf{C}}}=\left[\begin{array}{cc}
\mathfrak{R}\{\mathbf{C}\} & -\mathfrak{I}\{\mathbf{C}\}\\
\mathfrak{I}\{\mathbf{C}\} & \mathfrak{R}\{\mathbf{C}\}
\end{array}\right]$ for an arbitrary matrix $\mathbf{C}$. 

Similarly, the sensing SINR can be reformulated as 
\begin{equation}
\gamma_{s}=a/b, \label{eq:SINR_ref}
\end{equation}
where $\left[\begin{array}{c}
a\\
0
\end{array}\right]=\mathbf{\hat{d}}^{T}\mathbf{\hat{G}}\hat{\mathbf{F}}\mathbf{\hat{v}}\hat{\mathbf{v}}^{T}\mathbf{\hat{F}}^{T}\mathbf{\hat{G}}^{T}\left[\begin{array}{c}
\mathfrak{R}\{\mathbf{d}\}\\
\mathfrak{I}\{\mathbf{d}\}
\end{array}\right]$ and $\left[\begin{array}{c}
b\\
0
\end{array}\right]=\mathbf{\hat{d}}^{T}\hat{\mathbf{B}}\left[\begin{array}{c}
\mathfrak{R}\{\mathbf{d}\}\\
\mathfrak{I}\{\mathbf{d}\}
\end{array}\right]$. Also, the real and imaginary part of $\mathbf{B}$ can be obtained as
%
\begin{align}
\left[\begin{array}{c}
\mathfrak{R}\{\mathbf{B}\}\\
\mathfrak{I}\{\mathbf{B}\}
\end{array}\right] & \!=\!\mathbf{\hat{G}}\hat{\mathbf{F}}\sum_{k=1}^{K}\hat{\mathbf{W}}_{k}\mathbf{\hat{W}}_{k}^{T}\mathbf{\hat{F}}^{T}\left[\begin{array}{c}
\mathfrak{R}\{\mathbf{G}\}^{T}\\
-\mathfrak{I}\{\mathbf{G}\}^{T}
\end{array}\right]+\sigma_{z}^{2}\left[\begin{array}{c}
\mathbf{I}_{N_{r}}\\
\mathbf{0}_{N_{r}}
\end{array}\right]\label{eq:Bp}
\end{align}


Moreover, it should be mentioned that the matrix inversion in \eqref{eq:RX_xomb}
is implemented by the following identity. 
\begin{cor}
If $\mathbf{N}$ is the inverse of $\mathbf{M}$ (i.e., $\mathbf{N}=\mathbf{M}^{-1}$),
then we have:
\begin{align}
\mathfrak{R}\{\mathbf{N}\} & =(\mathfrak{R}\{\mathbf{M}\}+\mathfrak{I}\{\mathbf{M}\}\mathfrak{R}\{\mathbf{M}\}^{-1}\mathfrak{I}\{\mathbf{M}\})^{-1},\\
\mathfrak{I}\{\mathbf{N}\} & =-\mathfrak{R}\{\mathbf{M}\}^{-1}\mathfrak{I}\{\mathbf{M}\}\mathfrak{R}\{\mathbf{N}\}.
\end{align}
\end{cor}

Utilizing the expression \eqref{eq:Rk_ref} and \eqref{eq:SINR_ref},
as well as the constraints \eqref{eq:SINRmin} and \eqref{eq:User_space},
the loss function of the considered system can be formulated as 
\begin{equation}
f=-\sum_{k=1}^{K}R_{C,k}+\sum_{k=1}^{K}\nu_{d,k}(\text{d}_{\min}-\text{d}_{k})^{+}+\nu_{s}(\gamma_{0}-\gamma_{s})^{+},
\end{equation}
where $\text{d}_{k}=\underset{1\le b_{1}\neq b_{2}\le N_{k}}{\min}\;||\mathbf{\tilde{u}}_{k,b_{1}}-\mathbf{\tilde{u}}_{k,b_{2}}||$
is the smallest inter MA distance for use $k,$ $\nu_{d,k}$ is the
penalty parameter if $\text{d}_{k}$ does not satisfy the constraint
\eqref{eq:User_space}, and $\nu_{s}$ is the penalty parameter if
the SINR for target sensing violets the constraint \eqref{eq:SINRmin}. 

\vspace{-0.7em}
\section{Simulation Results}

In this section, we present the sum-rate of the considered system
using the proposed neural network by using Monte Carlo simulations.
As a benchmark, we consider a scheme that differs from the proposed
one in that it has fixed PAs, and the users' are equipped with fixed
antennas instead of MAs. This schemes is denoted as \emph{Fix-Ant},
and in it, the positions of the transmit PAs satisfy $y_{t,n}=(n-1)D_{t}/(N_{t}-1)$
and the user's MAs are placed in an ULA parallel to the $y$-axis
so that $\tilde{y}_{k,b}^{u}=(b-1)D_{k}/(N_{k}-1)$ for \textbf{$b\in\{1,\dots,N_{k}\}$}.

In the following simulation setup, the parameters are $f=5\thinspace\mathrm{GHz}$
($\lambda=6\,\mathrm{cm}$), $P_{\max}=1\,\mathrm{W}$, $N_{t}=6$,
$N_{r}=4$, $N_{k}=3$, $K=2$, $L=2$, $D_{t}=10\,\mathrm{m}$, $D_{k}=15\,\mathrm{cm}$,
$d_{\min}=\lambda/2=3\,\mathrm{cm}$, $\kappa=2$, $n_{e}=1.4$, $\gamma_{0}=0.01$,
and $\sigma_{k}^{2}=\sigma_{z}^{2}=-120\,\mathrm{dB}$. The midpoint
of the receive ULA is located at $(20\,\mathrm{m},20\,\mathrm{m},1\,\mathrm{m})$,
and the length of this array is $D_{r}=30\,\mathrm{cm}$. The position
of the first user is randomly selected from the space $[-5\,\mathrm{m},-3\,\mathrm{m}]\times[-5\,\mathrm{m},-3\,\mathrm{m}]\times[5\,\mathrm{m},10\,\mathrm{m}],$
while the position of the second user is selected from the space $[-5\,\mathrm{m},-3\,\mathrm{m}]\times[13\,\mathrm{m},15\,\mathrm{m}]\times[5\,\mathrm{m},10\,\mathrm{m}]$.
Similarly, the target is randomly located in the space $[12\,\mathrm{m},15\,\mathrm{m}]\times[0\,\mathrm{m},5\,\mathrm{m}]\times[1\,\mathrm{m},5\,\mathrm{m}]$.
All scatterers are randomly distributed over the space $[-2\,\mathrm{m},0]\times[3\,\mathrm{m},7\,\mathrm{m}]\times[0,3\,\mathrm{m}]$.
The proposed scheme is optimized using the stochastic gradient descent
(SGD) method together with the Adam optimizer. The used training set
consists of 50000 samples, while the test set has 10000 samples. The
learning rate is $\text{5}\times10^{-4}$ and the specified batch
size is 150 samples. The penalty parameters are $\nu_{s}=1000$ and
$\nu_{d,k}=100$ (for $k=1,2$). 
\begin{figure}[t]
\centering{}\includegraphics[scale=0.7]{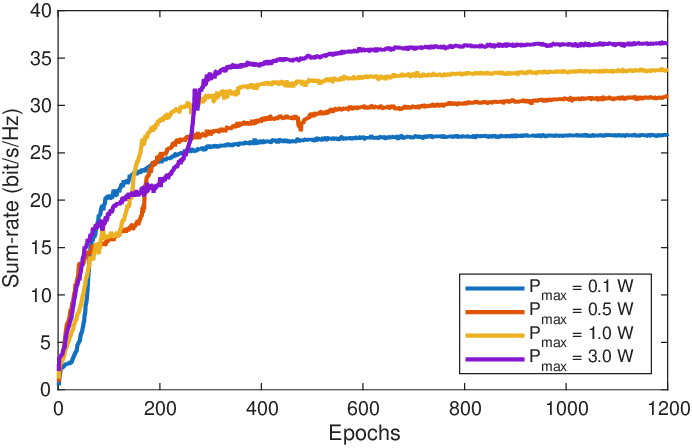}\caption{Convergence of the proposed neural network. \label{fig:Convergence}}
\end{figure}

In Fig. \ref{fig:Convergence}, we show the convergence of the proposed
neural network for different transmit power. In general, the sum-rate
does not always increase monotonically with the number of epochs,
and sometimes it can even drop sharply. This is caused by the change
of the batch samples in each epoch of the SGD method. Moreover, it
can be observed that the proposed network reaches 80\,\% of its convergent
rate in about 200 epochs, regardless of the value of the maximum transmit
power. Although the required number of epochs for network convergence
is not necessarily proportional to the transmit power, the lowest
number of epochs is needed when $P_{\max}=0.1\,\mathrm{W}$. 
\begin{figure}[t]
\centering{}\includegraphics[scale=0.7]{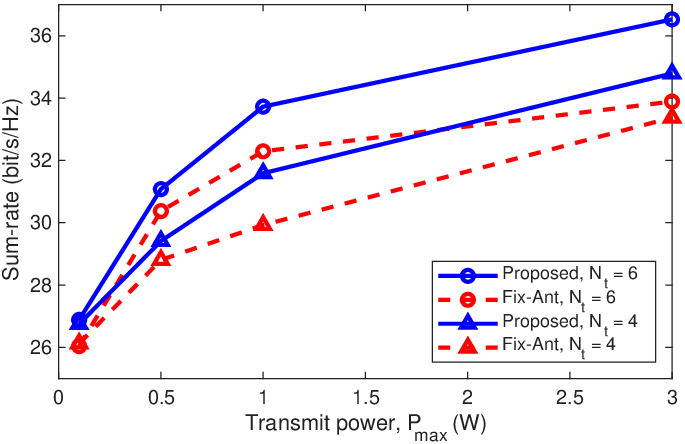}\caption{Sum-rate versus the maximum transmit power.\label{fig:Rate-vs-pow}}
\end{figure}

In Fig. \ref{fig:Rate-vs-pow}, we present the sum-rate for the proposed
neural network versus the maximum transmit power. The rates for both
the proposed and \emph{Fix-Ant} schemes increase approximately logarithmically
with the transmit power. As expected, the proposed scheme provides
a higher sum-rate because of its ability to optimize the positions of
the transmit PAs and receive users' MAs. Similarly as in \cite{perovic2025optimal},
this performance advantage is gradually enlarged with the transmit
power, although for $N_t=4$, it becomes almost constant for larger transmit powers. Hence, we can conclude that use of MAs and PAs is especially
beneficial in the middle to high transmit power range.

\begin{figure}[t]
\centering{}\includegraphics[width=8.85cm
]{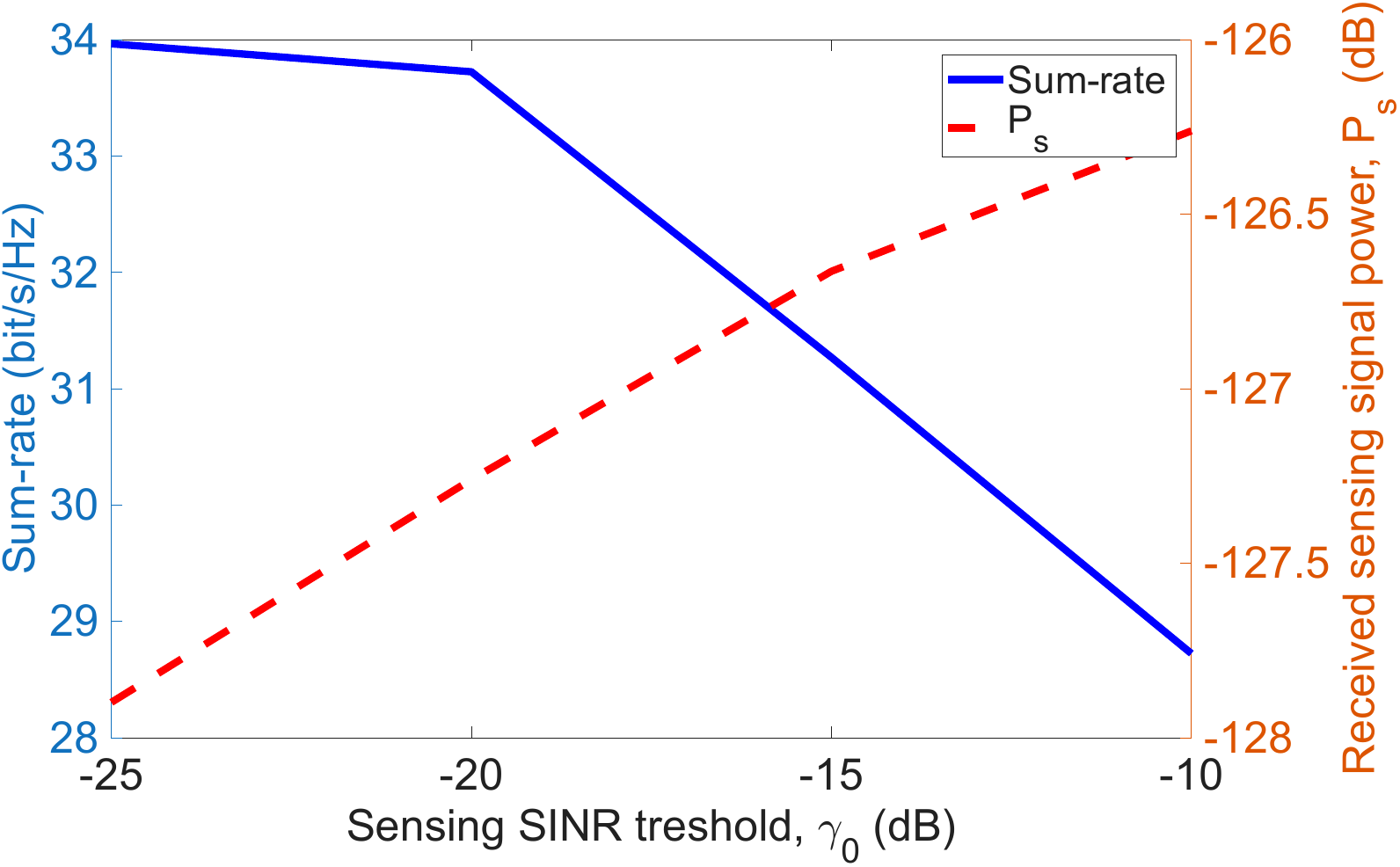}\caption{Sum-rate and sensing signal power vs. sensing SINR threshold.\label{fig:Rate-Ps}}
\end{figure}

In Fig. \ref{fig:Rate-Ps}, we show the sum-rate and the received sensing signal
power $P_{s}$ versus the sensing SINR threshold. As in \cite{perovic2025optimal}, the
sum-rate is decreased, while on the other hand the received sensing signal power
is increased with the SINR threshold.  
Furthermore, we observe that the communication rate is generally more sensitive to changes in the sensing SINR threshold than the received sensing signal power considered in this work.

\vspace{-0.4em}
\balance
\section{Conclusions}
\vspace{-0.2em}
In this paper, we studied the optimization of the sum-rate in an ISAC
system equipped with the transmit PAs and users' MAs. We derived the
closed-form solution for the optimal sensing receive combiner. In
addition, we provided a DL network that optimizes the positions of
the transmit PAs, the positions of the users' MAs, the communication
precoding matrices, and the sensing transmit beamformer in the considered
ISAC system. To train this network, we formulated the loss function
that consists of the objective function, as well as the appropriate
penalty terms related to the constraints for the PAs' and the MAs' positions.
Simulation results demonstrated that the performance advantage of
the proposed scheme in comparison to a benchmark scheme equipped with
only fixed antennas is gradually increased with the maximum transmit
power, and that the communication features of the considered
system were slightly more affected by the change of the sensing SINR
threshold than the sensing ones.

\vspace{-0.3em}
\bibliographystyle{IEEEtran}
\bibliography{IEEEabrv,IEEEexample,ML_PA_MA_NF_V2}

\end{document}